# Crystallite-size Dependent Harmonic Magneto-electricity in SmFeO$_3$

Pooja Sahlot and Anand Mohan Awasthi*

*UGC-DAE Consortium for Scientific Research, University Campus,
Khandwa Road, Indore- 452 001, India*

*amawasthi@csr.res.in

## Abstract

First- and second-harmonic dielectric susceptibilities are maidenly studied on Samarium Orthoferrite of mesoscopic/500 nm and nanoscopic/55 nm grainsizes. Magneto-electrically coupled to the antiferromagnetic and spin-reorientation transitions, fundamental and harmonic dielectricity consistently reflect the global/local polarization effects of crystallite-size dependent electrical orderings. Bulk and incipient ferroelectricity respectively in nanoscopic and mesoscopic crystallites concur the higher-temperature antiferromagnetic ordering ($T_N$ ~670 K). Upon the spin-reorientation transition at lower-temperature ($T_{SR}$ ~470 K), re-entrant relaxor state in the nano-crystallites and bulk-like/temperature-windowed ferroelectricity in the meso-crystallites emerge. In the nano-crystallites, magneto-electric signature of interfacial spins' de-pinning ($T_{SP}$ ~540 K) is exclusively revealed by the scaled-harmonics.

**Introduction**

RFeO$_3$ oxides are the perovskites with rich properties and functional applicability in spintronics, magneto-electric memory, solid oxide fuel cells, and many others [1,2,3]. Perovskites and related systems with multiple magnetic phases and respective transitions have shown interesting magneto-electric properties in the literature [4,5]. In RFeO$_3$ family, samarium orthoferrite (SmFeO$_3$; SFO) with single-phase ambient multiferroicity is of great interest [6,7]. Here, G-type antiferromagnetic ordering of Fe$^{3+}$ ions with weak ferromagnetism along the *c*-axis occurs below $T_N$ =670 K [8,9]. Upon further cooling, the SFO system shows reorientation of Fe$^{3+}$ spins from the *c*- axis to *a*- axis below $T_{SR}$ =470 K [10]. Fe-Sm interactions are reported to play a crucial role in this magneto-strictive transition [11]. Significant displacement of Sm-ion in the octahedra introduces (improper) polarization in SFO [12], presenting good candidate multiferroic for study.

Chaturvedi et. al. [9,13] have carried out detailed particle-size dependent study on SFO. The group witnessed magneto-dielectric coupling about $T_N$ in SFO nanoparticles (~55 nm) [13], which has been attributed to exchange striction and significant intrinsic surface stress of the nanoparticle-shell. Raman measurements in the studied system evidenced spin-phonon coupling across $T_N$ and $T_{SR}$. In SFO with meso-sized grains (~500 nm), qualitative change in the ac-conductivity mechanism across $T_{SR}$ was reported via its Jonscher power-law analysis [9]. Interestingly, the strength and type of magneto-electric (ME) couplings about $T_{SR}$ and $T_N$ were observed to depend on the grain size. In this regard, we expect the nature and magnitude of ME-induced polarizations to be crystallite-size dependent, providing important functional implications. To precisely characterize the global/local polarizations and thus identify the exact electrical orderings across the temperatures of interest, we present the first upgraded dielectric study in SFO; involving the harmonic susceptibility investigations. These were demonstrated in our previous report [14], to manifest profound signatures of unusual polarizations, which decipher several variant electrically-ordered states in different perovskite-related systems.

Here, we have performed dielectric fundamental, first-, and second-harmonics study on two SFO specimens- with grains of mesoscale ($m$-SFO ~500 nm) and nanoscale size ($n$-SFO ~55 nm). Samarium Orthoferrite (SmFeO$_3$) has orthorhombic symmetry (Pnma), with distorted FeO$_6$ octahedra [9]. Magnetic study on the specimens found their antiferromagnetic transition (Fe$^{3+}$ spins) at $T_N$ =670 K and spin-reorientation at 470 K (disorder-broadened; ±10 K)--where weak-ferromagnetic moment associated with the Fe-spins reorients from the $c$-axis to the $a$-axis [9]. Chaturvedi et. al. reported dielectric properties for $n$-SFO (avg. particle size ≈55±5 nm) with non-dispersive $\varepsilon'$-peaks at $T_N$, wherein no sharp anomaly could be observed for $m$-SFO specimen [9]. From first- and second-harmonic dielectric measurements, here we present clear and profound ME-coupling driven polarization-characters across both $T_N$ and $T_{SR}$, with well-distinguished magneto-harmonic effects in $n$-SFO and $m$-SFO specimens—as crystallite-size dependent non-linear dipolar-response. Further, novel magneto-electric feature across the interfacial-spins' depinning temperature ($T_{SR}$< $T_{SP}$ <$T_N$) is exclusively revealed from scaled-harmonic susceptibility.

## Dielectric- Fundamental and Harmonics Study on *m*-SFO (500 nm grain-size)

Fundamental dielectric measurements were performed on the *m*-SFO specimen, where fig. 1 presents temperature dependence (300-750 K) of dielectric constant $\varepsilon'(T)$ in the frequency range of 100 Hz to 100 kHz. $\varepsilon'(T)$ isochrones depict no sharp feature at $T_N$ (670 K); upon cooling there is steep drop in loss-tangent (fig.1 (b)) and a local-plateau in $\varepsilon'(T)$ (fig.1 (a)) around ~600 K. Across the spin reorientation transition ($T_{SR}$ ~470K) however, clear non-dispersive peak-anomaly shows in both dielectric-constant and loss-tangent (fig.1 (a, b)). Here, low-valued $\varepsilon'$ and losses (tan$\delta$ < 1) reflect intrinsic nature of magneto-dielectric anomaly, indicating a bulk-like polarization.

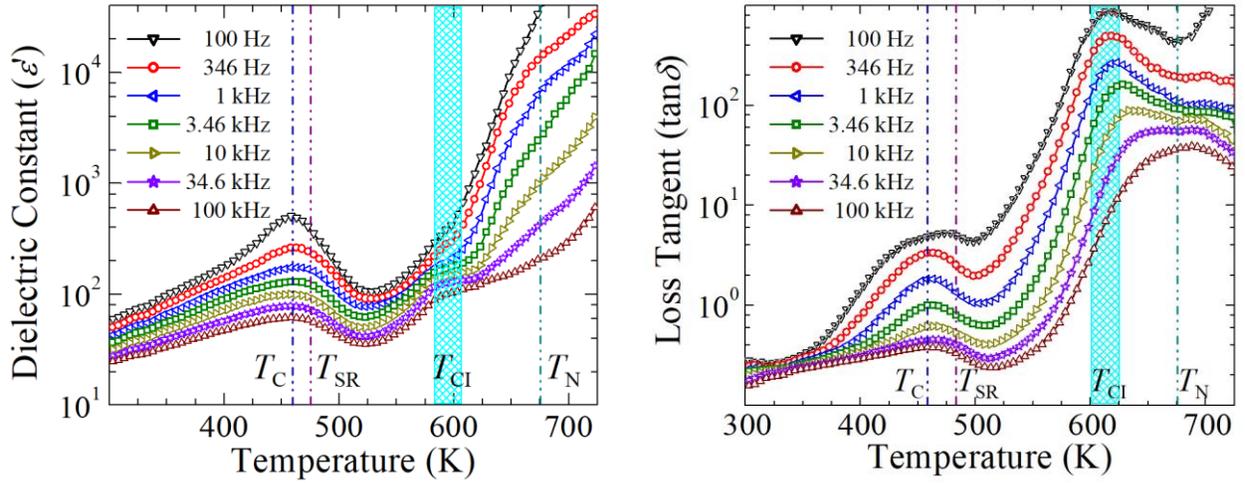

Figure 1. Temperature dependence of (a) dielectric constant ($\varepsilon'$) and (b) loss-tangent for *m*-SFO.

For further rigorous investigation, non-linear dielectric measurements of first- and second-order harmonics were performed across 300-750 K over (instrument-limited) frequency range of 100 Hz to 750 Hz. First and second harmonics directly probe the polarization (*P*) in the system [15];

$$\varepsilon'_2 = -3\varepsilon_0^2 B P \chi'^3 \qquad (1)$$

$$\varepsilon'_3 = -(1 - 18\varepsilon_0 B P^2 \chi')\varepsilon_0^3 B \chi'^4 \qquad (2)$$

Here, B is the positive coefficient of bipolar polarization-term ($P^2$) in the free energy. Second-harmonic reflects the combined effect of bilinear-term in polarization and fundamental susceptibility ($\chi'$), and competition between the two determines the nature of the second-harmonic signal. Sets of the first and second harmonics have been analyzed in [14], depicting the polarization phenomenon with temperature dependence in several perovskite-related systems.

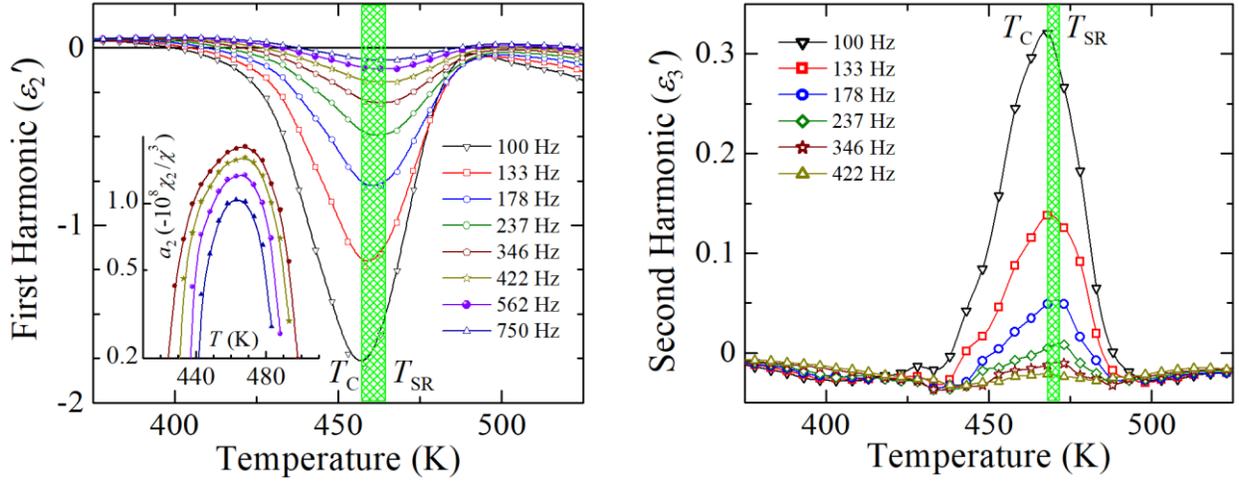

Figure 2. Temperature dependent first- and second-harmonic signals for *m*-SFO.

First- and second-harmonic susceptibilities measured for *m*-SFO system show no clear anomaly exactly at $T_N$. Figure 2 presents the harmonic signals- $\varepsilon_2'(T)$ and $\varepsilon_3'(T)$ across $T_{SR}$. The negative-peaks in $\varepsilon_2'$ indicate a net polarization *P* (eq.1) in the system, consistent with the non-dispersive peaks in the fundamental signals (fig.1). Concurrent positive peaks in $\varepsilon_3'$ further affirm bulk-like ferroelectricity in the mesoscopic grains, establishing a 'Type-II' multiferroicity, Mesoscopic-grains with 'canted-AFM Néel-domains' accommodate extended Néel-domain-wall, having *c*- to *a*-axis modulating spins. Advocated rather early by Kimura [16] and Mostovoy [17], symmetry-analyzed by Zvezdin [18], and lately reported e.g., in $Li_{0.05}Ti_{0.02}Ni_{0.93}O$ [19] and $Ca_2FeCoO_5$ [20] among others, such (short-ranged spiral/cycloidal) spin-modulations induce---via inverse Dzyaloshinskii-Moriya (ID-M) spin-orbit interaction---ME-polarization throughout the Néel-domain-wall interstitial-matrix. Effectively, this mimics 'bulk-ferroelectricity'--- albeit existent only over the disorder-broadened $T_{SR}$-window, wherein the activated-process of *c*- to *a*-axis spin-reorientation occurs. The scaled first-harmonic ($a_2(T) = -\varepsilon_2'/\chi'^3 \propto P$) amply evidences this in fig.2 (a)-inset, undergoing an order of magnitude rise & fall over $\Delta T_{SR} \sim 80$ K.

Scaled $a_2(T)$ at higher temperatures (fig. 3) depicts non-dispersive *peak* anomaly at ~600 K. Upon cooling below $T_N$, losses drop sharply across this temperature (fig.1 (b)), with concurrent local-plateau in the dielectric constant (fig.1 (a)). These consistent features signify the *incipient* ferroelectric nature of the *m*-SFO system below $T_N$; having only *dynamic* dipole-correlations, with their most prominent demarcating signature in the *scaled* first-harmonic signal.

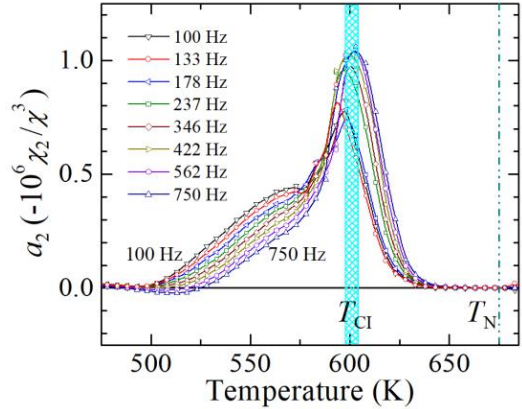

Figure 3. Scaled first-harmonic signal $a_2(T)$, presenting anomaly below $T_N$ across ~600 K.

## Dielectric- Fundamental and Harmonics Study on *n*-SFO (55 nm grain-size)

Figure 4 presents $\varepsilon'(T)$ isochrones for the *n*-SFO specimen. In consistency with literature [9], $\varepsilon'(T)$ shows frequency-independent peak-anomaly in the dielectric constant across $T_N$ (fig.4 (b)). Concurrently, sharp decrease in the losses is observed, with tan$\delta$ dropping below 1 (fig.5), indicating intrinsic dielectric contribution below $T_N$. This affirms that for nano-metric grains, equivalence of crystallite-size & electrical correlation length well conserves the system's magneto-electricity across $T_N$. At the meso-scale, disorder comes into effect and degrades the ME-coupling at high-temperatures. Upon further cooling below $T_{SR}$, the bulk-like FE-correlations (well-emergent in the *m*-SFO specimen) get less-pronounced/smeared-out in the case of the *n*-SFO specimen (fig.4(a)). In consistency with the literature [13], dispersive peaks below $T_{SR}$ in fig. 4(a) and fig. 5 correspond to the relaxor-FE state, re-entrant upon spin-reorientation, from the parent bulk ferroelectric state (realized upon the antiferromagnetic transition at $T_N$).

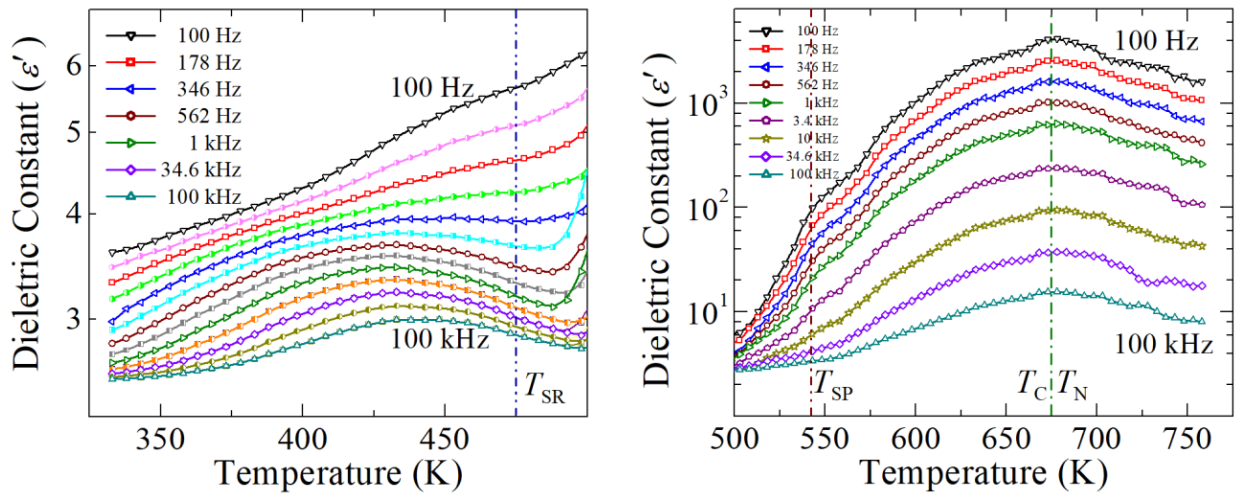

Figure 4. Temperature dependent dielectric constant for *n*-SFO across (a) $T_{SR}$ and (b) $T_N$.

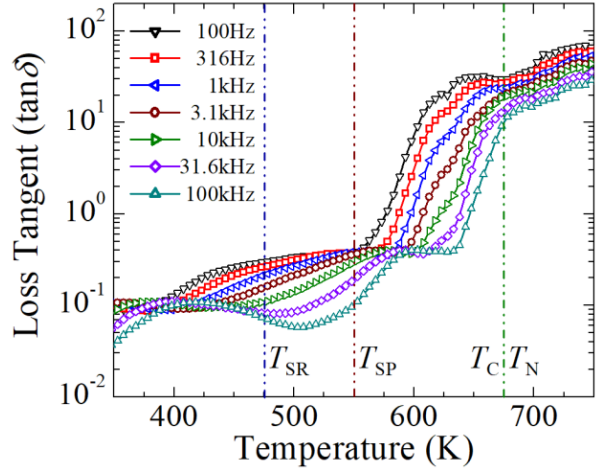

Figure 5. Temperature dependence of loss-tangent for $n$-SFO

Across $T_N$ and $T_{SR}$, first- and second-harmonic measurements are performed on $n$-SFO specimen. Figure 6(a) presents negative-peaks in $\varepsilon'_2$ harmonic-signal across the AFM transition in $n$-SFO. Here, the negative anomaly in $\varepsilon'_2$ with the positive $\varepsilon'_3$-signal across $T_N$ (fig.6 (b)) affirm an ME-coupling induced robust polarization state. Moreover, scaled $a_2(T)$ as a metric of the polarization $P$, rises by five orders of magnitude upon cooling below the $T_N/T_C$ (fig. 6(a) inset). These consistent behaviors establish clear Type-II multiferroicity here. In the $n$-SFO specimen, although frequency-dispersive peaks in dielectric constant are observed across spin reorientation region, no anomaly-signature is present in the harmonic-signals there; both $\varepsilon'_2$- and $\varepsilon'_3$-signals tend to zero. Reentrant relaxor-ferroelectric state here below $T_{SR}$ signifies Type-I multiferroicity, upon spin-reorientation.

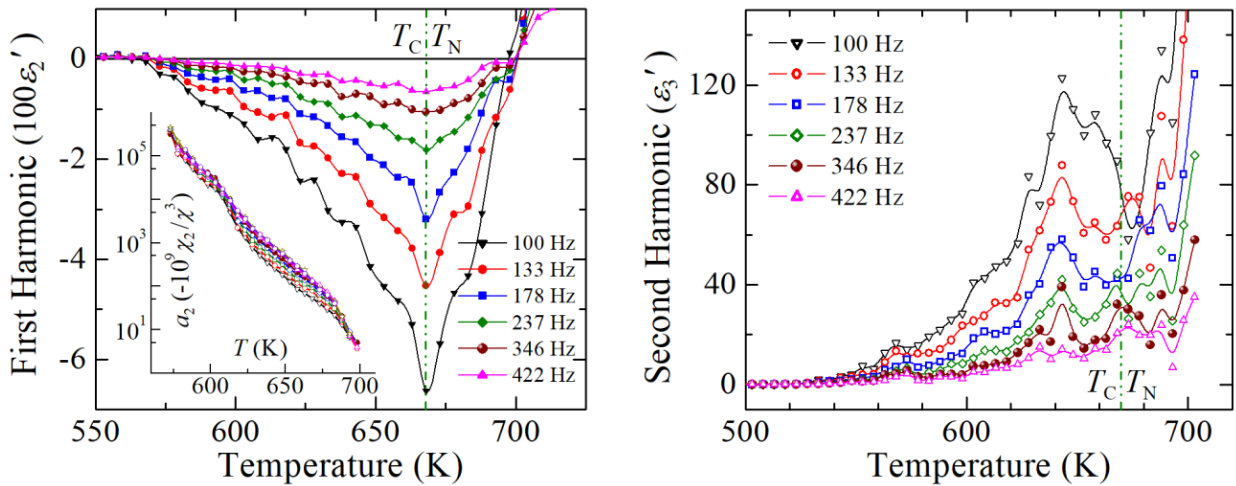

Figure 6. Temperature dependent signals of (a) first-harmonic and (b) second-harmonic in $n$-SFO.

Manifestation of hyper-polarizations $(\varepsilon'_2, \varepsilon'_3 \neq 0)$ requires broken electrical-isotropy (non-vanishing internal global/local fields). In the *n*-SFO specimen, the individual nanoparticles as Néel-ferrimagentic mono-domains feature unidirectional WFM-spins throughout their entirety. The absence of Néel-wall matter to host modulating spins deprives the system of an ID-M-like mechanism for realizing domain-wall polarization, which could elicit non-linear/harmonic magneto-dielectric response over the $T_{SR}$-regime. Of course, the magnetic frustration---borne of spin-orientations distributed amongst the nanoparticulate Néel-domains, with its allied magneto-electric disorder---retreats the range of dipole-correlations to sub-particle-size length-scale, thereby suppressing the parent bulk ferroelectricity into a re-entrant relaxor-FE state.

Further, scaled susceptibilities evaluated as $a_2 = -\varepsilon'_2/\chi'^3$ and $a_3 = \varepsilon'_3/\chi'^4$ shown in fig. 7 present the temperature dependence of these scaled-harmonic signals. A pronounced anomaly is present in both the scaled susceptibilities across ~545 K, which is not resolvable in the bare/measured harmonic-signals, and is rather extinct in the fundamental signals. In the *n*-SFO specimen, a magnetic anomaly at 550 K [9] is attributed to the thermal activation of the pinned interfacial spins in the nano-grains. Clear signature of the concurrent anomaly in scaled harmonic-susceptibilities here exclusively manifests a novel magneto-electric effect, of the subtle magnetic phenomenon in the SFO-nanoparticles, for the first time.

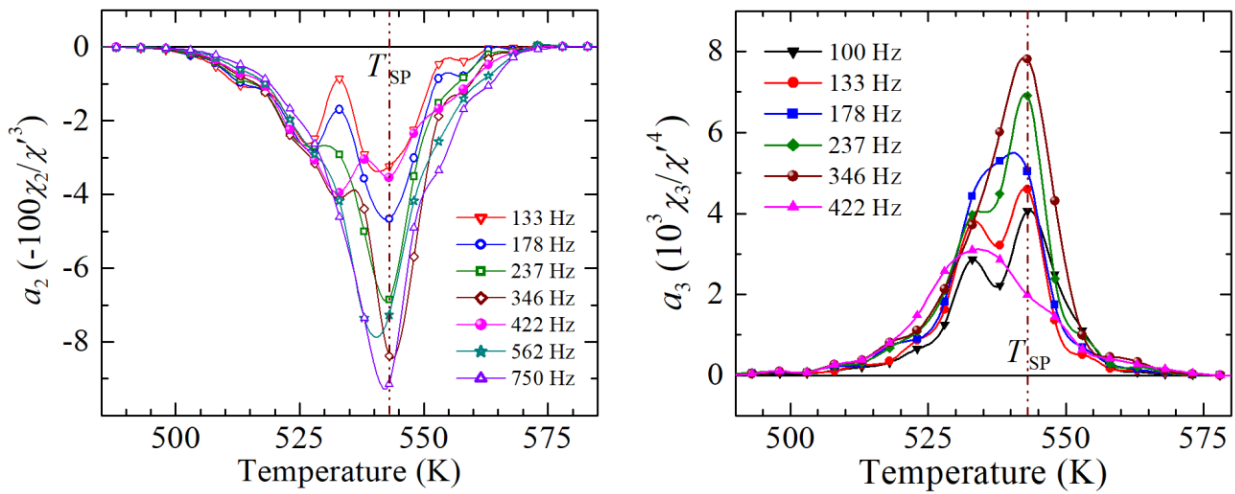

Figure 7. Temperature dependent signals of scaled parameters; (a) $a_2(T)$ and (b) $a_3(T)$ in *n*-SFO.

## Conclusions

Maiden first- and second-harmonic non-linear dielectric study on Samarium Orthoferrite presents novel crystallite-size dependent magneto-electric effects, obtained on mesoscopic and nanoscopic grain-sized specimen. In mesoscopic specimen, magneto-striction driven bulk-like polarization (Type-II multiferroicity) across spin-reorientation transition-window ($\Delta T_{SR}$) is established, whereas around 600 K (below $T_N$), clear signature of incipient ferroelectricity is affirmed by the scaled harmonic susceptibility. In nanoscopic specimen, exchange-striction and intrinsic-surface-stress driven robust ferroelectricity at $T_N$ ($\equiv T_C$; Type-II MF) is established from all-non-dispersive peak-anomalies in fundamental, first-harmonic (negative) and second-harmonic (positive) signals. Here, upon magneto-strictive spin-reorientation, the emergence of re-entrant relaxor-FE state (Type-I MF) and the absence of harmonic signals, are both traceable to the nanometric particles as Néel-ferrimagnet mono-domains---precluding the Néel-domain-wall matter and ensuing magneto-electric disorder. From scaled harmonic-signals in the nanoscopic specimen, novel magneto-electric effect marking thermal activation of interfacial-pinned spins is witnessed for the first time. Our findings reiterate the importance of harmonic dielectric investigations as exclusive upgradation tools in probing intricate/disordered polarizations and identifying complex electrical phases, which are incompletely characterized by the conventional transport parameters.

## Acknowledgments

Authors are grateful to Sulabha Kulkarni, Smita Chaturvedi, and their group-members at IISER-Pune, India, for consenting their Samarium Orthoferrite samples (*n*-SFO and *m*-SFO) for the study.